# INTERRELATIONAL MODEL FOR UNDERSTANDING CHATBOT ACCEPTANCE ACROSS RETAIL SECTORS

Diptish Dey*, and Debarati Bhaumik*

March 09, 2022

## Abstract

*Despite the rising interest in chatbots, deployment has been slow in the retail sector. In the absence of comparative cross sector research on the user acceptance of chatbots in retail, we present a model and a research framework that proposes customer and chatbot antecedents using trust and customer satisfaction as relationship mediators and word of mouth and expectation of continuity as relationship outcomes. In determining our framework, we assimilate constructs from different models and theories overarching user experience with chatbots, technology acceptance and relationship marketing and propose a selection of 11 constructs as antecedents. Furthermore, we suggest retail sectors as one of our 4 moderators. Eventually, we provide insight into our current activities that is expected to identify which factors impact relationship outcomes to which extent across different retail sectors.*

## 1 Introduction

The consumers and shoppers of today are increasingly high-tech. Empowered by AI-enabled chatbots, websites, social media and messaging services are increasingly making use of these technologies to manage customer experience. Business Insider (2020) predicts the chatbot market to experience strong growth soon. Despite the strong reference to the near future, it is worthwhile to reconcile that the adoption of technology for example in retailing has a relatively long history (Watson, 2011) and its impact on customer-seller exchanges has been summarized by Hagberg, et al. (2016).

Recent study (Taylor, et al., 2020) conducted at a global level indicate a satisfaction level of 62% with chatbots found on a website. The study further points out that 74% of organizations surveyed considered "conversational assistants as a key enabler of the company's business and customer engagement strategy". Despite this mutual willingness, the actual deployment levels across sectors have been lagging; in the consumer products & retail sector 23% of organizations deploy a chatbot.

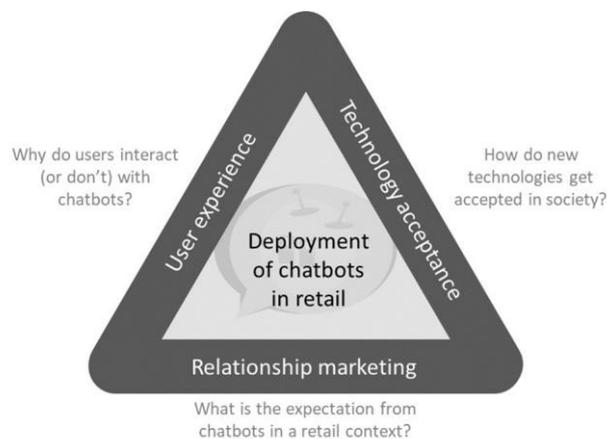

*Figure 1: Interrelational model*

*Amsterdam University of Applied Sciences, Fraijlemaborg 133, 1102 CV Amsterdam, The Netherlands

e-mail: d.dey2@hva.nl

In this paper we approach the problem of lagging deployment of chatbots across retail sectors by first presenting our findings, based on literature study, along three different aspects: user experience, technology acceptance and relationship marketing. This is illustrated in figure 1. The first aspect (user experience) revolves around the user interaction with chatbots: why do users interact with chatbots, what motivates them to use chatbots and what discourages them from doing so? The second aspect (technology acceptance) relates to how new technologies get accepted in society. The third aspect (relationship marketing) formalises the expectation that people (users and non-users) have from a chatbot in a retail sector. The following sections present a summary of these aspects based on literature study.

In a real world, interplay of these three aspects is expected to drive chatbot deployment and usage; albeit these aspects may not be exhaustive nor mutually exclusive. We limit ourselves to the combined role of these three aspects and propose a framework that provides us insights into the following research questions: (1) which factors drive chatbot relationship outcomes in retail? (2) how do these factors differ in their impact on chatbot relationship outcomes across different retail sectors? Subsequently empirical research into these questions combined with multilevel analysis and structured equation modelling will drive conclusions on the behavorial differences that chatbots must demonstrate across retail sectors. In our opinion it is this differentiation that should eventually lead to improved user experience and higher technology acceptance of chatbots in retail.

## 2  User experience with chatbots

The motivation of people to interact with a chatbot has been a topic of research analyzed through different theories. The social response theory (Nass and Moon 2000; Nass et al. 1994) views the interaction between a human being and a computer as fundamentally social: the perception that computers are social actors occurs automatically and unconsciously. Since then, multiple studies have provided evidence on how social rules apply to anthropomorphically designed computers. More recently, this theory has been applied to chatbots (Adam, et al., 2020) to understand the many unsatisfactory encounters that users have with chatbots.

In addition to the social response theory, research has focused on the pragmatic quality of chatbot dialogues (Li et al., 2020 and Shevat, 2017). Some parameters that users rate as important within such dialogues include *conversational intelligence*, *retention of conversational context* (Jain, et al., 2017), *message interactivity* (Go and Sundar, 2019), *conversational flow* (Gnewuch et al., 2018 and Skuve et al., 2019), *conversational repair* (Ashktorab, et al., 2019), and *adaptability to user type* (Ruane, et al., 2020). In most research, the extraction of information from chatbot dialogues is highly automated in nature, making use of text mining (Akhtar et al., 2019) and sentiment analysis (Feine et al., 2019). Følstad and Taylor (2021) present a more up-to-date framework for qualitative analysis of chatbot dialogues to provide insight into key drivers of user experience.

Uses and gratification (U&G) theory, which helps to understand the gratifications that people obtain through proactive media consumption (Rubin, 1983), has been used to understand the gratification that users obtain when interacting with chatbots (Brandtzaeg and Følstad, 2017). Cheng and Jiang (2020) have logically clustered gratifications into four categories: Utilitarian, Hedonic, Technology, and Social. These categories have been used to research gratifications, and its impact on levels of user satisfaction, obtained from chatbots serving different consumer brands. Utilitarian gratification refers to the individuals' *utility needs* from a chatbot (Papacharissi & Mendelson, 2011); hedonic gratification refers to the *emotional gratification* that users can receive or the fun of using a chatbot (Brandtzaeg and Følstad, 2017); *technology gratification* refers to the ability of chatbots to be more precise and efficient as compared to their human counterparts (Sundar & Kim, 2019); social gratification as suggested by Araujo (2018) refers to the adoption of chatbots due to their *social presence* and the existence of an artificial being that responds to users.

# 3    Technology acceptance

The topic of individual acceptance of technology has been widely researched with different streams of research focusing on different aspects. For instance, Davis et al. (1989) lay their emphasis upon intention or usage as the dependent variable when researching individual acceptance of technology. Whereas Leonard-Barton and Deschamps (1988) have focused on implementation success at the organizational level. The former, i.e., intention of usage has been widely used in information systems research as a vital predictor of technology acceptance (Taylor and Todd, 1995).

Some models and theories of individual acceptance of technology available in literature are: Theory of Reasoned Action (TRA), Technology Acceptance Model (TAM), Motivational Model (MM) and Innovation Diffusion Theory (IDT). TRA has its roots in social psychology and its core constructs are *attitude toward behaviour* and *subjective norm* (Fishbein and Ajzen, 1975). The former specifies a user's "positive or negative feelings (evaluative affect) about performing the target behavior", whereas the latter stipulates the user's "perception that most people who are important to him think he should or should not perform the behavior in question".

TAM, which grew in popularity with the advent of information systems and technologies, includes *perceived usefulness*, *perceived ease of use* and *subjective norm* as core constructs; the latter was adopted from the TRA model (Venkatesh and Davis, 2000). Perceived usefulness indicates the extent to which a user believes that using a technology would enable him or her to do a better job. Perceived ease of use reflects the effort required by a user to use the technology.

MM relies on general motivation theory as an explanation for users' behavior (Vallerand, 1997) and its core constructs include *extrinsic motivation* and *intrinsic motivation*. Davis et al. (1989) defines extrinsic motivation as the desire of users to perform an activity "because it is perceived to be instrumental in achieving valued outcomes that are distinct from the activity itself, such as improved job performance, pay, or promotions". On the other hand, intrinsic motivation is defined as as the desire of users to perform an activity "for no apparent reinforcement other than the process of performing the activity per se".

IDT, which is anchored in sociology (Rogers, 1995) could in principle be applied to any type of innovation. In applying IDT to information systems Moore and Benbasat, (1991) defined a set of constructs such as *relative advantage*, *ease of use*, *image*, *visibility*, *compatibility*, *results demonstrability,* and *voluntariness of use*. Relative advantage indicates how good an innovation is as compared to the going concern. Ease of use indicates how difficult or easy an innovation is in usage. Image indicates how a social image of a user is enhanced by the usage of the innovation. Visibility reflects how visible the usage of a system is within an organization of network. Compatibilty as defined by Moore and Benbasat (1991), reflects "the degree to which an innovation is perceived as being consistent with the existing values, needs, and past experiences of potential adopters". The extent to which the relative advantage of the technology is tangible to people beyond the user is expressed as results demonstrability. Voluntariness of use reflects how voluntarily (free-willed) a user can use the system.

# 4    Relationship marketing

Summarizing the factors influencing the effectiveness of relationship marketing Palmatier et al., 2006 presents a meta-analytic framework, which encompasses customer-focused, seller-focused & dyadic antecedents, relational mediators, moderators and customer-focused, seller-focused & dyadic outcomes. These antecedents, mediators and outcomes are in turn composed of constructs.

Constructs within customer-focused antecedents include *relationship benefits* and *dependence on seller*. Relationship benefits could be functional or social in nature (Reynolds and Beatty, 1999). Functional benefits include *timesaving*, *convenience*, *tailored advice,* and *better purchase decisions*. An example of a social benefit is *companionship*. Dependence on seller refers to the customer's *perceived value of seller provided resources*.

Seller-focused antecedents include *relationship investment* and *seller expertise*, with the latter reflecting the *knowledge*, *experience,* and *overall competence* of the seller (Crosby, et al., 1990). Although a salesperson's attributes such as similarity and expertise influence sales effectiveness directly, only expertise has been found to influence the long-term sales relationship. Relationship investment refers to the time, effort and *resources* that sellers invest in building stronger relationships.

Dyadic antecedents are *communication*, *similarity*, *relationship duration*, *interaction frequency* and *conflict*. Communication builds stronger relationships within a formal or informal dialogue by helping resolve disputes, align goals, managing expectations and uncover new value creating opportunities (Morgan and Hunt, 1994). Communication that is frequent and of high quality (relevant, timely and reliable) results in greater trust between the customer and the seller. Doney and Cannon (1997) elaborates on how similarity, relationship duration and interaction frequency contribute to greater trust. As conflict increases, the customer is less likely to have confidence in the long-term orientation of the seller or to invest in building or maintaining a relationship; thus, conflict should negatively influence the customer's trust in and commitment toward the seller (Anderson & Weitz, 1992).

Relational mediators comprise of *trust*, *commitment*, *relationship satisfaction* and *relationship quality*. Trust could be defined as confidence in an exchange partner's reliability and integrity. A trading partner is trustworthy if and only if one is willing to take actions that otherwise would involve risk. "Though it certainly would be appropriate to incorporate 'stated willingness' in a measure of trust, willingness is unnecessary or redundant in its definition" (Morgan and Hunt, 1994). Commitment is defined as an enduring desire to maintain a valued relationship (Moorman et al., 1992). Because trust increases the extent to which users engage in risky exchanges, trust is expected to increase the likelihood that users will become committed in relationships. Satisfaction with the relationship is regarded as an important outcome of buyer-seller relationships. Relationship quality is viewed as a cumulative effect over the course of a relationship as opposed to satisfaction only, which is more transactional in nature (De Wulff et al., 2001).

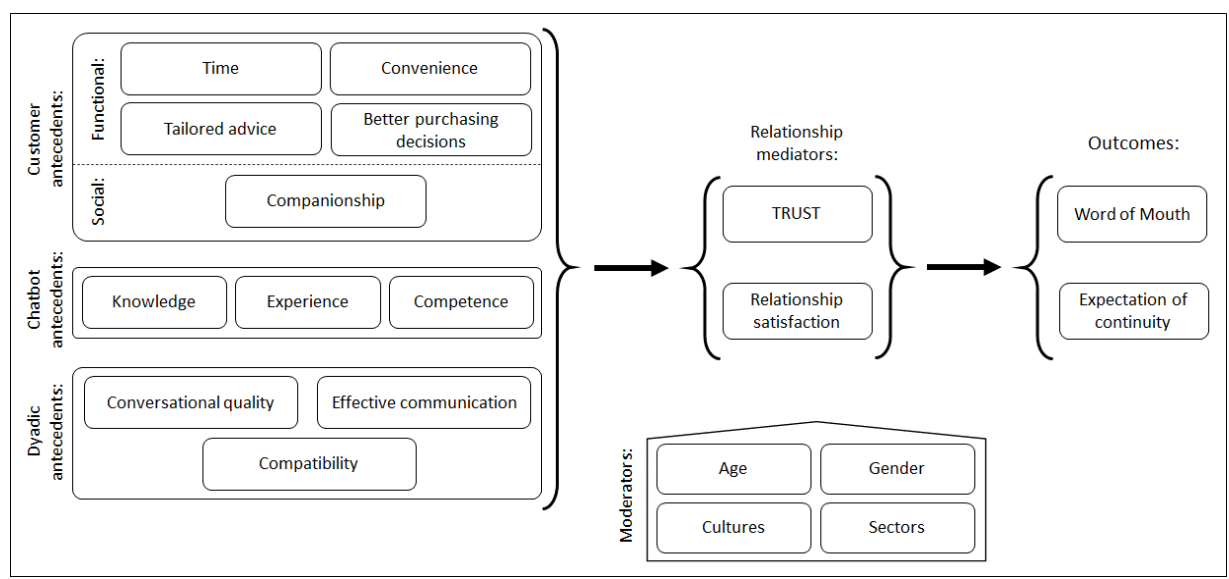

*Figure 2: Framework for research*

Like antecedents, relationship outcomes include customer-focused, seller-focused and dyadic outcomes. Constructs within customer-focused outcomes are *expectation of continuity*, *word of mouth* and *customer loyalty*. Expectation of continuity reflects the likelihood of continued purchases. Word of mouth reflects the likelihood op positive seller referrals. Although many measures of loyalty exist in literature, higher loyalty is often reflected in higher share of wallet (Keiningham, et al., 2011). Examples of seller-fcossed outcomes include *performance enhancements* such as *sales growth*, *profit performance* and increased *share of wallet* (Reynolds and Beatty, 1999). Dyadic antecedents include *coordinated and joint actions* between the customer and the seller (Morgan and Hunt, 1994).

Moderators influence the relationships within the framework to varying degrees. Palamtier et al., (2006) include *service versus product-based exchanges*, *channel versus direct exchanges*, *business versus consumer markets,* and *individual versus organizational relationships* as moderators in their framework. These moderators differ based on the context of the study. Venkatesh and Davis (2000) for an online context have used moderators such as *gender*, *age*, *experience,* and *voluntariness of use*. Verma et al., (2015) have reduced the number of constructs of the meta-analytic framework and considered only those constructs which are relevant when applying the framework to online retailing.

# 5 Research framework

In conducting the literature study of the three aspects as presented in the interrelational model (figure 1), we processed multiple models and theories discussed in sections 2, 3 and 4. This resulted in the assimilation of construcst, which are indicated in italics in the above sections. Subsequently we analysed these constructs to arrive at the research framework presented in figure 2. In doing so, we have merged constructs that are overlapping to a great extent (e.g., *utility needs* and *functional benefits*). We dropped constructs (e.g., *results demonstrability*) that have limited relevance with our scope of researching user acceptance of chatbots across different retail sectors. We grouped the antecedents into customer, chatbot and dyadic. This is because we assume in our research that the seller relationship will be delivered through a chatbot instead of a human seller.

We have omitted *commitment* and *relationship quality* as relationship mediators from the framework. This is because of two reasons: first and foremost, *trust* and *relationship satisfaction* are to a large extent precursor to *commitment* and *relationship quality* respectively (Morgan and Hunt, 1994); secondly, our research is focused on user acceptance, which is a relatively preliminary stage in the innovation decision process (Rogers, 1995). This rationality has been furthermore applied to the outcomes, which has resulted in the omission of advanced outcomes such as *customer loyalty* and *performance enhancements* and has emphasized *word of mouth* and *expectation of continuity* as more relevant relationship outcomes for our research.

In proposing moderators, we have not only adopted moderators from the literature study presented in the preceeding sections but also preferred those that have a strong purpose in our research. The former has resulted in the inclusion of *age*, *gender*, and *cultures* as moderators, whereas the latter has strongly influenced our inclusion of *sectors* as a moderator. Sectors imply retail sectors, a comparative study of which is a consequence of our research and is expected to result in unique insights.

# 6 Ongoing activities

To validate the framework presented in figure 2, we are conducting an empirical research. Hypotheses linking antecedents to mediators and mediators to outcomes have been formulated. Upon conclusion of the empirical research, we intend to use classical statistics, testing for interactions and derive conclusions for each antecedent and mediator pair. In addition to the pairwise analysis, we aim to integrate the analyses into a single multivariate model.

Our aim is to construct a multilevel model with the moderators as shown in figure 2 (age, gender, cultures, and retail sectors) as our random factors. This will enable us to capture variation within the regression coefficients at the level of the moderators. The purpose of the multilevel model is descriptive and causal as opposed to predictive modelling. Furthermore, the multilevel model will be non-nested, given that the moderators are overlapping and non-hierarchical. Eventually, it is expected to help us comprehend how the impact of antecedents on the mediators and the outcomes vary across the moderators, especially across the retail sectors.